\documentclass[showpacs,preprintnumbers,amsmath,amssymb]{revtex4}
\usepackage{graphicx}% Include figure files
\usepackage{dcolumn}% Align table columns on decimal point
\usepackage{bm}% bold math
\begin{document}
\title{On the reformulation of Thomas-Fermi to make it compatible to the Planck-scale}

% Force line breaks with \\
\author{Himangshu Barman, Anisur Rahaman}

\email{anisur.rahman@saha.ac.in;
 manisurn@gmail.com}\affiliation{Hooghly Mohsin College, Chinsurah, Hooghly - 712101,
West Bengal, India}
\author{Sohan Kumar Jha}
\affiliation{Chandernagore College, Chandernagore, Hooghly, West
Bengal, India}

\date{\today}% It is always \today, today,
             %  but any date may be explicitly specified

\begin{abstract}
\begin{center}
Abstract
\end{center}
Thomas-Fermi model is considered here to make it cogent to capture
the Planck-scale effect with the use of a generalization of
uncertainty relation. Here generalization contains both linear and
quadratic terms of momentum. We first reformulate the Thomas-Fermi
model for the non-relativistic case. It has been shown that it can
also be reformulated for taking into account the relativistic
effect. Dialectic screening for the non-relativistic cases has
been studied and the expression of screening length has been found
out explicitly.
\end{abstract}
\maketitle

\section{Introduction}
Over the decades, there has been a huge interest in the study of
predicting the probable behavior of the physical system in the
vicinity of Plank-scale. Three are different ways of studying the
probable behavior of physical systems in the vicinity of
Planck-scale through some well-designed formulation which has a
very close link with the well-developed theory (string theory)
believed to be suitable to the Plank-scale regime. The
generalization of uncertainty relation is one of the most
important and well-accepted frameworks in this respect. It
acquired a great deal of attention when it has been found to be
concomitant with the string theory, loop quantum gravity, and
non-commutativity of space-time as well, at the conceptual level
\cite{KEMPF, DAMITQM, MAGG1QM, MAGG2QM}. The generalization of
uncertainty relation is made in various ways however the
generalization associated with minimum length and minimum length
along with the maximum momentum are the two basic criteria of the
generalization since the one-dimensional object namely {\it
string} which is supposed to appear at the Plank scale would
appear with the length of order of Planck-length.

There are a heap of literature on these two types of generalized
uncertainty relation in the different branches of physics.
Although generalization of uncertainty generally called {\it
generalized uncertainly principle} $(GUP)^1$ \footnotetext[1]{We
will prefer to use generalization of uncertainty relation in place
of principle to extend our due respect to the celebrated
Heisenberg uncertainty principle that brought a breakthrough in
fundamental development of physics.} with quadratic term in
momentum initially showed the way to introduce that concept of
minimum length, the concept of minimum length along with the
maximum momentum was developed later through linear-quadratic GUP
and these two are equally potent to incorporate the Planck-scale
correction. The articles \cite{DAMITQM, MAGG1QM, MAGG2QM, MAGG3QM,
PPQM, BPMQM, KEMPF, PPHGUP, PPHGUP1} offer lucid and elaborate
discussion on how the concept of minimum length is correlated to
the well-developed theory of Plank-scale (string theory).

The formal development of statistical physics to make it amenable
to the Planck-scale is carried out in the articles \cite{ALIST,
PPSTAT, CHENGPS, HOMACOS, AFALIPS, VEGPSCOS}. In condensed matter
physics the the generalization of uncertainty relation is used for
the same purpose in the articles \cite{ASVC3, ASVC4, ASVC5, ASVQG,
HOMA1, HOMA2}. In black-hole physics also we find the extensive
use of different types of deformed uncertainty algebra in the
articles \cite{GRAV1, GRAV2, KNOZ1GRAV, KNOZ2GRAV, SUNANDAGRAV,
MIRGRAV, MIRML, MIREFIELD, KIM1BH, KIM2BH, VEGBH} to incorporate
Planck-scale effect. The Quark-Gluon Plasma physics too has been
extended with the GUP framework in the article \cite{VEGQGP,
ALIQGP, NAGQGP, ARQGP} for the same purpose.

The recent work of Shebabi {\it etal.} \cite{HOMA1, HOMA2} on the
Thomas-Fermi(Thomas \cite{THOM}, Fermi \cite{FERMI}) are the
fascinating extension of generalized uncertainty relation having
quadratic momentum dependence. Thomas-Fermi (TF) model is a
many-particle statistical model that came up as a heuristic
semi-classical method to describe the electrostatic potential and
the charge densities in large atoms, metals, and in astrophysical
objects such as neutron stars \cite{STAR, STAR1}. Here the
electrons are considered to behave like ideal gas that obeys the
Fermi-Dirac statistics. In its original formulation, the effect of
exchange forces is not taken into account, and the temperature of
the system is taken as $T=O$. Dirac has extended the theory to
include the effects of exchange forces. Relativistic corrections
over it were developed by Vallarta and Rosen \cite{VALL} by
replacing the non-relativistic electron kinetic energy term in the
Thomas-Fermi formulation by its relativistic counterpart. On the
other hand, the thermal effects on the Thomas-Fermi formulation
were considered by Marshak and Bethe \cite{BETHE} and Feynman,
Metropolis, and Teller \cite{FEYN}, among others the thermal
effects are as important as the relativistic effects if the
temperature are of the order of $10^7~ {}^0K$ or more. In its
original formulation, the effect of exchange forces was not taken
into account, and the temperature of the system is taken as $T=O$.
Dirac has extended the theory to include the effects of exchange
forces \cite{DIR}. The extension of Shebabi {\it etal}. would be
useful to capture the Plank-scale physics not only in the
solid-state physics but also in the astrophysics. In the article,
the author has studied the TF model with a special type of GUP. So
a natural extension of this model is the application of another
GUP on it and study the outcome of it. We have already mentioned
that there is another class of GUP that is associated with the
minimum length along with the maximum momentum. The introduction
of linear-quadratic GUP and its interesting extension in different
branches of physics are found in the articles \cite{VEGLQ1,
VEGLQ2, VEGLQ3, VEGLQ4, VEGLQ5, PPLQ1, PPLQ2, PPLQ3, KNOZLQTF}.
And in many cases, the extension of the same model in different
GUP has come up since the basic foundation of the deformations
differs in a distinct manner. So, naturally, the correction
followed are also distinctly different.

The TF model is so popular and useful in providing the different
aspects of the charged many-particle system that extension of this
model with other deformed Heisenberg algebra would be instructive.
In this respect, it would be of worth investigation of
Thomas-Fermi model with the generalized Heisenberg algebra having
linear and quadratic quadratic term of momentum. In this article,
we, therefore, make an attempt to carry out the formulation of the
TF model with the the generalized Heisenberg algebra having both
linear and quadratic term of momentum. We should mention here that
in the article \cite{KNOZLQTF}, the TF model has been attempted to
solve with this type of generalized algebra, however the approach
we will follow here is different. In \cite{KNOZLQTF}, the
correction is introduced  at the operator level within the
Hamiltonian and a perturbation technique is used to evaluate the
energy of the system. But in our extension correction will be
incorporated through the density of state-level as has been found
in \cite{HOMA1, HOMA2} following the formalism \cite{SHIV}.

The article is organized as follows. In Sec.I, a general
discussion over Heisenberg algebra with linear and quadratic term
in momentum is given. Sec II is devoted withy the formulation of
the Thomas-Fermi model with Linear-Quadratic generalization to
make it cogent to capture the Planck-scale phenomena. In Sec. III,
the screening effect is studied with this reformulated model. Sec.
IV is an extension to include a relativistic effect in the present
situation. A brief summary and discussion are given in Sec. V.

\section{General description of Heisenberg algebra with the
linear and quadratic term in momentum}
The celebrated Heisenberg algebra is given by
\begin{equation}
 [x, p]= i\hbar\label{HCOM},
\end{equation}
and the uncertainty relation from which the above algebra
generates is
\begin{equation}
\Delta x \Delta p \ge \frac{\hbar}{2}. \label{QCOM}
\end{equation}
Generalized Heisenberg algebra with quadratic term in momentum
reads
\begin{equation}
[x, p]= i\hbar(1+ \beta^2 p^2). \label{COMC}
\end{equation}
However for generalization of Heisenberg with linear and quadratic
term in momentum algebra has the the following form:
\begin{equation}
[x, p]= i\hbar(1-2\alpha p + 4\alpha^2 p^2). \label{COM}
\end{equation}
The uncertainty relation corresponding to the algebra (\ref{COM})
is
\begin{equation}
\Delta x, \Delta p \ge \frac{\hbar}{2}[(1+
\frac{\alpha}{\sqrt{<p^2>}}+4\alpha^2)\Delta p^2
+4\alpha^2<p>^2-2\alpha\sqrt{<p>^2}].
\end{equation}
Here $\alpha = \frac{\alpha_0}{m_p  c}=\alpha_0\frac{l_p}{\hbar}$,
where $m_p c^2 = 10^{19}GeV$,  and the Planck length $l_p =
10^{-35}m$. Unlike the quadratic generalization , concept of
maximum momentum along with the minimum length is admissible here,
and here lies the fundamental difference between these two.  The
minimum length and maximum momentum admissible to this deformed
algebra respectively are
\begin{equation} \delta x\ge\delta x_{min} \approx \alpha_0 l_p,
\delta p\le\delta p_{min} \approx \frac{m_0 c}{\alpha_0 }.
\end{equation}
For $D$ dimension the equation (\ref{COM}) gets generalized to the
following
\begin{equation}
[x_i, p_j]= i\hbar[\delta_{ij}-\alpha(\delta_{ij}p +
\frac{p_ip_j}{p})+ \alpha^2(\delta_{ij} p^2 +3p_ip_j)].
\label{COMG}
\end{equation}
Note that the equation (\ref{COMG}) is satisfied by the following
representation of position and momentum respectively:
\begin{eqnarray}
x_i&=&x_{0i},\nonumber\\
p_i&=&p_{0i}(1-\alpha p + 2\alpha^2p^2).
\end{eqnarray}
The invariant phase space volume for this deformation entails the
following modification
\begin{equation}
[D\mu]= \frac{d^3x d^3p}{J},
\end{equation}
where $J$ is the jacobin for transformation that reads
\begin{equation}
J^{-1}= (2\pi)^3[1-\alpha p+ (\frac{2}{D+1}+ \frac{1}{2})\alpha^2
 p^2]^{D+1}\label{JAC}
\end{equation}
for the deformed uncertainty relation considered here.  These are
the necessary input for this generalized uncertainty relation
having  linear and quadratic term of momentum to deal with it. We,
are therefore, in a position to apply it to the model we have
considered for extension. To be precise we will consider the
Finite temperature TF model and pursue the non-relativistic case
to start with in the framework  to which we now turn.
\section{Finite temperature Thomas-Fermi model with a generalized
uncertainty relation}
 Thomas-Fermi model model is formulated with the  consideration
that a many electron system is equivalent to a gas of fermions
obeying Fermi-Dirac statistics and these electrons are occupying
the phase space uniformly with one spin up and one spin down
electron per unit cell having volume $h^3$. Therefor, the density
of electron is given
\begin{equation}
n(r)= \frac{N}{V}= g\frac{2}{h^{3}}\int_{}^{}f(E)d^{3}p,
\label{DEN}
\end{equation}
when position and momenta satisfy the usual Heisenberg algebra.
Here $N$ represents the total number and $V=\int d^3 x$ . The
Fermi-Dirac distribution function is denoted by $f(E)$:
\begin{equation}
F(E)=\frac{1}{e^{\frac{(E-\mu)}{KT}}+1},\label{DENF}
\end{equation}
and $g$ represents the degeneracy. In this situation $g=2$. So the
total number of electron can be obtained  by summing or
integrating over the energy $E$ as applicable according to the
nature of the problem. According to the Thomas-Fermi model the
screened coulomb potential is given by
\begin{equation}
\nabla^2 \Phi= 4\pi e(\eta-\eta_0) - 4\pi q \delta(r),
\end{equation}
where where $\eta$ and $\eta_{0}$ are the number of particles in
the excited and ground state respectively. We, therefore, have
\begin{equation}
\nabla^2 \phi= 4\pi e(I-I_0) - 4\pi q \delta(r),
\end{equation}
where
\begin{equation}
I=g\frac{2}{h^{3}}\int_{}^{}\frac{1}{J}\frac{4\pi p^2
dp}{exp(\frac{\frac{p^2}{2m}-\tilde{\mu}}{KT}+1)}
\end{equation}
here $J$ represents the jacobian for transformation as given in
(\ref{JAC}), and
\begin{equation}
I(0) = I_{(\phi=0)}.
\end{equation}
and $\tilde{\mu}=\mu+e\phi$. Here $\phi$ represents electrostatic
potential, $\mu$ is the chemical potential, $T$ stands for
temperature and $K$ is the Boltzmann constant. It is a general
practice to set $K=1$ without any loss of generality. In the limit
of large volume, it is reasonable to use integral over all phase
space keeping jacobian within $J$ since the correction enterers
through that.

The generalized uncertainty which we are going to use to study the
finite temperature Thomas-Fermi model to be potent to acquire
Planck-scale effect is
\begin{equation}
[x, p]= i\hbar(1-\eta p + \eta^2 p^2) \label{COMU}
\end{equation}
It is convenient to we use $2\alpha = \eta$ to get rid of factors
of $2$ or its multiples in different parts of the computation. So
the phase space volume in presence of this generalized Heisenberg
algebra.
\begin{eqnarray}
D\mu &=& \frac{d^3xd^3p}{h^3}(1-\eta p + \eta^2 p^2)^{-4}\nonumber
\\
&=&\frac{d^3xd^3p}{h^3}(1+4\eta p +6\eta^{2}p^{2}
-20\eta^{3}p^{3}+10\eta^{4}p^{4} + ........).
\end{eqnarray}
Here $\eta\ll 1$. So the density of states retaining the terms up
to order $\eta^2$ is given by
\begin{equation}
D\mu \approx \frac{d^3xd^3p}{h^3}(1-4\eta p + 6\eta^2 p^2).
\end{equation}
Using the above expression of density states, the particle density
per unit volume $n=\frac{N}{V}$ is obtained which is given by
\begin{eqnarray}
n&=&\frac{2\pi}{h^{3}}(2mT)^{\frac{3}{2}}[\,\int_{0}^{\infty}\frac{y^{\frac{1}{2}}dy}
{e^{y-\frac{\xi}{T}}+1}
+4\eta(2mT)^{\frac{1}{2}}\int_{0}^{\infty}\frac{ydy}{e^{y-\frac{\xi}{T}}+1}
+6\eta^{2}(2mT)\int_{0}^{\infty}\frac{y^{\frac{3}{2}}dy}{e^{y-\frac{\xi}{T}}+1}
\nonumber\\
&-&20\eta^{3}(2mT)^{\frac{3}{2}}\int_{0}^{\infty}\frac{y^{2}dy}{e^{y-\frac{\xi}{T}}+1}
+10\eta^{4}(2mT)^{2}\int_{0}^{\infty}\frac{y^{\frac{5}{2}}dy}{e^{y-\frac{\xi}{T}}+1}]\,
\label{NF}
\end{eqnarray}
In the above expression we make the substation
$y=\frac{\epsilon}{T}$ and $\epsilon = \frac{p^{2}}{2m}$. The
equation (\ref{NF}) contains fermi integral
\begin{equation}
f_{\nu}(\varsigma)=\frac{1}{\Gamma(\nu)}\int_{0}^{\infty}\frac{y^{\nu-1}dy}{e^{y-\varsigma}+1}
\label{DEF}.
\end{equation}
where $\Gamma(\nu)$ is the Euler Gamma function. In weakly
non-degenerate case $|\varsigma|\gg 1$. With this limit the
equation(\ref{DEF}) has the expansion
\begin{equation}
f_{\nu}(\varsigma)=\frac{\varsigma^{\nu}}{\Gamma(\nu +
1)}[1+\nu(\nu+1)\frac{\pi^{2}}{6\varsigma^{2}}+.......],
\end{equation}
So the Thomas-Fermi density of a non-relativistic particle in the
presence of the deformed Heienberg algebra is found out to be
\begin{eqnarray}
n&=&\frac{\sqrt{2}}{\pi^{2}\hbar^{3}}(me)^{\frac{3}{2}}(\phi+\frac{\mu}{e})^{\frac{3}{2}}[\{\frac{1}{3}
+\frac{\pi^{2}T^{2}}{24e^{2}(\phi+\frac{\mu}{e})^{2}}\}
+\eta(2me)^{\frac{1}{2}}(\phi+\frac{\mu}{e})^{\frac{1}{2}}\{1+\frac{\pi^{2}T^{2}}{3e^{2}
(\phi+\frac{\mu}{e})^{2}}\}\nonumber\\
&+&\frac{6}{5}\eta^{2}(2me)(\phi+\frac{\mu}{e})\{1+\frac{5\pi^{2}T^{2}}{8e^{2}(\phi+\frac{\mu}{e})^{2}}\}
-\frac{10}{3}\eta^{3}(2me)^{\frac{3}{2}}(\phi+\frac{\mu}{e})^{\frac{3}{2}}\{1+\frac{\pi^{2}T^{2}}{e^{2}
(\phi+\frac{\mu}{e})^{2}}\}\nonumber\\
&+&\frac{10}{7}\eta^{4}(2me)^{2}(\phi+\frac{\mu}{e})^{2}\{1+\frac{35\pi^{2}T^{2}}{24e^{2}
(\phi+\frac{\mu}{e})^{2}}\}].\label{TFD}
\end{eqnarray}
The use  Poisson's equation along with the TF density (\ref{TFD})
enables us to get generalized TF equation:
\begin{eqnarray}
\nabla^{2}\phi &=& 4\pi en\simeq 4\pi
e\frac{\sqrt{2}}{\pi^{2}\hbar^{3}}(me)^{\frac{3}{2}}
(\phi+\frac{\mu}{e})^{\frac{3}{2}}[\{\frac{1}{3}
+\frac{\pi^{2}T^{2}}{24e^{2}(\phi+\frac{\mu}{e})^{2}}\}
 \nonumber\\
&+&\eta(2me)^{\frac{1}{2}}(\phi+\frac{\mu}{e})^{\frac{1}{2}}\{1+\frac{\pi^{2}T^{2}}{3e^{2}
(\phi+\frac{\mu}{e})^{2}}\}
+\frac{6}{5}\eta^{2}(2me)(\phi+\frac{\mu}{e})\{1+\frac{5\pi^{2}T^{2}}{8e^{2}
(\phi+\frac{\mu}{e})^{2}}\}
\nonumber \\
&-&\frac{10}{3}\eta^{3}(2me)^{\frac{3}{2}}(\phi+\frac{\mu}{e})^{\frac{3}{2}}\{1+\frac{\pi^{2}T^{2}}{e^{2}
(\phi+\frac{\mu}{e})^{2}}\}
+\frac{10}{7}\eta^{4}(2me)^{2}(\phi+\frac{\mu}{e})^{2}\{1+\frac{35\pi^{2}T^{2}}{24e^{2}
(\phi+\frac{\mu}{e})^{2}}\}] \label{POLQ}
\end{eqnarray}
The terms containing $\eta$ represent the correction due to the
use of deformed Heisenberg algebra. Note that the above equation
lands onto the usual Thomas-Fermi equation if we set  $\eta=0$.
The equation (\ref{POLQ}) is the generalized form of the
Thomas-Fermi equation  with linear-quadratic generalization which
will be equally useful to capture the Plank-scale effect.

\section{The dielectric screening process in presence generalized uncertainty relation}
As an application we will now turn towards the investigation of
the impacts of this alternative deformed Heisenberg algebra on
dielectric screening process. To this end we assume a uniform gas
of electrons having charge density $-n_{0}e$ is superimposed on a
lattice of shielded nuclei with charge density $n_{0}e$ and place
a positive point charge  $Q$ as a test charge in this charged sea
with a coulomb potential $\phi$ to study the screaming effect in
this situation. It is straightforward to obtain the screening
 potential since the screening
potential would be a solution of Poison's equation in this
particular environment.
\begin{equation}
\nabla^{2}\phi=4\pi e(n-n_{0})-4\pi Q\delta(r). \label{POIS}
\end{equation}
Our objective is to incorporate the Planck-scale effect through
the specified generalized uncertainty. So it is to bear in mind
that the density that has to be used is the the modified  TF
density to have a solution of the equation (\ref{POIS}) with
$g=2$. So what we  have is
\begin{eqnarray}
\nabla^{2}\phi &=& 2\frac{4me^{2}\sqrt{2me}}{\pi
\hbar^{3}}\{(\phi+\frac{\mu}{e})^{\frac{3}{2}}[\{\frac{1}{3}+\frac{\pi^{2}T^{2}}{24e^{2}
(\phi+\frac{\mu}{e})^{2}}\}
+\eta(2me)^{\frac{1}{2}}(\phi+\frac{\mu}{e})^{\frac{1}{2}}\{1+\frac{\pi^{2}T^{2}}{3e^{2}
(\phi+\frac{\mu}{e})^{2}}\}\nonumber\\
&+&
\frac{6}{5}\eta^{2}(2me)(\phi+\frac{\mu}{e})\{1+\frac{5\pi^{2}T^{2}}{8e^{2}(\phi+\frac{\mu}{e})^{2}}\}
-\frac{10}{3}\eta^{3}(2me)^{\frac{3}{2}}(\phi+\frac{\mu}{e})^{\frac{3}{2}}
\{1+\frac{\pi^{2}T^{2}}{e^{2}(\phi+\frac{\mu}{e})^{2}}\}\nonumber\\
&+&
\frac{10}{7}\eta^{4}(2me)^{2}(\phi+\frac{\mu}{e})^{2}\{1+\frac{35\pi^{2}T^{2}}{24e^{2}
(\phi+\frac{\mu}{e})^{2}}\}]\nonumber\\
&-& (\frac{\mu}{e})^{\frac{3}{2}}
[(\frac{1}{3}-\frac{\pi^{2}T^{2}}{24\mu^{2}})+\eta(2m\mu)^{\frac{1}{2}}(1+\frac{\pi^{2}T^{2}}{3\mu^{2}})
+\frac{6}{5}\eta^{2}(2m\mu)(1+\frac{5\pi^{2}T^{2}}{8\mu^{2}})\nonumber\\
&-&\frac{10}{3}\eta^{3}(2m\mu)^{\frac{3}{2}}(1+\frac{\pi^{2}T^{2}}{\mu^{2}})
+\frac{10}{7}\eta^{4}(2m\mu)^{2}(1+\frac{35\pi^{2}T^{2}}{24\mu^{2}})]\}-4\pi
Q\delta(r).
\end{eqnarray}
If we keep ourselves restricted with the linear response only
ignoring all nonlinear effect  it will allow us to consider
the$|\frac{e\phi}{\mu}|\ll 1$.  With this approximation the above
equation gets simplified into
\begin{equation}
\nabla^{2}\phi=\frac{3}{2}\phi(\frac{4\pi n_{0}e^{2}}{\mu})
[1-\frac{\pi^{2}T^{2}}{24\mu^{2}}+\frac{3m\pi^{2}T^{2}\eta^{2}}{2\mu}
-\frac{40\sqrt{2}}{3}\frac{m^{\frac{3}{2}}\pi^{2}T^{2}\eta^{3}}{\mu^{\frac{1}{2}}}+25(m\pi
T\eta^{2})^{2}]-4\pi Q\delta(r). \label{SCREEN}
\end{equation}
Note that temperature independent term is not considered here. The
equation (\ref{SCREEN}) can be casted in to following simplified
form
\begin{equation}
\nabla^{2}\phi=\frac{3}{2}\phi [\lambda_{F}^{(\eta)}]^{-2}-4\pi,
Q\delta(r)
\end{equation}
where $\phi$ is given by
\begin{equation}
\phi=\frac{Q}{r}e^{-\sqrt{\frac{3}{2}}(\frac{r}{\lambda_{F}^{(\eta)}})}.
\end{equation}
Here
$n_{0}=\frac{1}{3\pi^{2}}(\frac{2m\mu}{\hbar^{2}})^{\frac{3}{2}}$
and the screening length length is found out to be
\begin{equation}
[\lambda_{F}^{(\eta)}]^{-2}=\frac{4\pi
n_{0}e^{2}}{\mu}(1-\frac{\rho}{3}+6\sigma-40\tilde{\sigma}+25\zeta)
\end{equation}
where
$\rho=\frac{\pi^{2}T^{2}}{8\mu^{2}}$,~$\sigma=\frac{\pi^{2}T^{2}m\eta^{2}}
{4\mu}$,~$\tilde{\sigma}=\frac{\sqrt{2}}{3}\frac{m^{\frac{3}{2}}\pi^{2}T^{2}\eta^{3}}{\sqrt{\mu}}$
and $\zeta=(m\pi T)^{2}\eta^{4}$.
\section{Relativistic Thomas-Fermi Model with linear-quadratic generalization}
We are now going to generalized the TF model with linear-quadratic
GUP to incorporate the Planck-scale correction for relativistic
domain. The modified phase space volume which follows from
equations(\ref{JAC}), (\ref{DEN}) and (\ref{DENF}).
\begin{equation}
D\mu\simeq\frac{d^{3}x d^{3}p}{h^{3}}(1+4\eta p +6\eta^{2}p^{2}).
\end{equation}
retaining the terms  up to 2nd order in $\eta$. So the the TF
density in the relativistic regime is described by
\begin{equation}
n(r)=\frac{1}{\pi^{2}\hbar^{3}}\int_{0}^{\infty}\frac{p^{2}dp(1+4\eta
p
+6\eta^{2}p^{2})}{1+exp\{b[\sqrt{p^{2}c^{2}+m^{2}c^{4}}-mc^{2}-e(\phi
-\phi_{0})]\}},
\end{equation}
where $m$ is the mass of the particle, $b$ is just the inverse
temperature since $K=1$ is already set. However,  the exact
expression of $b$ ($b\equiv \frac{1}{KT}$), and $\mu = -e\phi_{0}$
is the chemical potential.

It would be useful to introduce a Juttner's transformation
\cite{JUT} at this stage since it will make the computation
tractable a lot. Explicitly it is
\begin{equation}
\frac{p}{mc}=sinh\theta.
\end{equation}
With this transformation the TF density reads
\begin{eqnarray}
n(r)&=&\frac{m^{3}c^{3}}{\pi^{2}\hbar^{3}}\int_{0}^{\infty}\frac{sinh^{2}\theta
cosh\theta d\theta(1+4\eta mc
sinh\theta+6\eta^{2}m^{2}c^{2}sinh^{2}\theta)}{1+exp\{b[\sqrt{m^{2}c^{4}sinh^{2}\theta+m^{2}c^{4}}-mc^{2}
-e(\phi-\phi_{0})]\}}\nonumber\\
&=&\frac{m^{3}c^{3}}{\pi^{2}\hbar^{3}}\int_{0}^{\infty}\frac{sinh^{2}\theta
cosh\theta d\theta(1+4\eta mc
sinh\theta+6\eta^{2}m^{2}c^{2}sinh^{2}\theta)}{1+\frac{1}{\Lambda}exp(bmc^{2}cosh\theta)}\nonumber\\
&=&\frac{m^{3}c^{3}}{\pi^{2}\hbar^{3}}\{\int_{0}^{\infty}\frac{sinh^{2}\theta
cosh\theta
d\theta}{1+\frac{1}{\Lambda}exp(bmc^{2}cosh\theta)}+4\eta
mc\int_{0}^{\infty}\frac{sinh^{3}\theta cosh\theta
d\theta}{1+\frac{1}{\Lambda}exp(bmc^{2}cosh\theta)}\nonumber\\
&+&6\eta^{2}m^{2}c^{2}\int_{0}^{\infty}\frac{sinh^{4} \theta
cosh\theta d\theta}{1+\frac{1}{\Lambda}exp(bmc^{2}cosh\theta)}\},
\label{RTFD}
\end{eqnarray}
where $\Lambda = e^{[b(\mu mc^{2}+e\phi(r))]}$. If we now adopt a
new variable  ${\omega}$ in the similar way has been used in
\cite{CHAN}
\begin{equation}
 \omega=bmc^{2}cosh\theta.
\end{equation}
the above equation (\ref{RTFD}) reduces to
\begin{eqnarray}
n(r)&=&\frac{m^{3}c^{3}}{\pi^{2}\hbar^{3}}\frac{1}{bmc^{2}}[\int_{0}^{\infty}\frac{
\frac{d\xi_{1}(\omega)}{d\omega}}
{{1+\frac{1}{\Lambda}e^{\omega}}}d\omega+4\eta
mc\int_{0}^{\infty}\frac{ \frac{d\xi_{3}(\omega)}{d\omega}}
{{1+\frac{1}{\Lambda}e^{\omega}}}d\omega
\nonumber \\
&+& 6\eta^{2}m^{2}c^{2}\int_{0}^{\infty}\frac{
\frac{d\xi_{2}(\omega)}{d\omega}}
{{1+\frac{1}{\Lambda}e^{\omega}}}d\omega], \label{RRTFD}
\end{eqnarray}
where
\begin{eqnarray}
\frac{d\xi_{1}}{d\omega}=sinh\theta cosh\theta \label{DP1},
\end{eqnarray}
\begin{eqnarray}
\frac{d\xi_{2}}{d\omega}=sinh^{2}\theta cosh\theta \label{DP2},
\end{eqnarray}
\begin{eqnarray}
\frac{d\xi_{3}}{d\omega}=sinh^{3}\theta cosh\theta \label{DP3}.
\end{eqnarray}
At this stage it would be useful to use the use the Sommerfeld
lemma \cite{SOM} which leads us to have the following simplified
from from the equation (\ref{RRTFD}):
\begin{eqnarray}
n(r)&=&
\frac{m^{3}c^{3}}{\pi^{2}\hbar^{3}}\frac{1}{bmc^{2}}[\xi_{1}\omega_{0})
+\frac{\pi^{2}}{6}\xi_{1}^{''}(\omega_{0})+.......\nonumber\\
&+&6\eta^{2}m^{2}c^{2}(\xi_{2}(\omega_{0})
+\frac{\pi^{2}}{6}\xi_{2}^{''}(\omega_{0})+........)\nonumber\\
&+& 4\eta mc(\xi_{3}(\omega_{0})
+\frac{\pi^{2}}{6}\xi_{3}^{''}(\omega_{0})+.......)],\label{FDEN}
\end{eqnarray}
where $\omega_0$ is function of $\Lambda$ which has the following
explicit expression
\begin{equation}
\omega_{0}=ln\Lambda=b[\mu+mc^{2}+e\phi(r)]=
bmc^{2}cosh\theta_{0}.
\end{equation}
To obtain the expression of $n(r)$ in a desired form  a new
variable $s$ is introduced which has the following definition
\begin{equation}
s=sinh\theta_{0}=[\frac{(\mu+mc^{2}+e\phi(r))}{m^{2}c^{4}}-1]^{\frac{1}{2}},
\end{equation}
the equations (\ref{DP1}), (\ref{DP2}) and (\ref{DP3}) in terms of
$s$ look
\begin{eqnarray}
\xi_{1}(\omega_{0})=\frac{bmc^{2}}{3}s^{3},{}\xi_{1}^{''}(\omega_{0})=\frac{1}{bmc^{2}}(\frac{2s^{2}+1}{s}),\label{SDP1},
\end{eqnarray}
\begin{eqnarray}
\xi_{3}(\omega_{0})=\frac{bmc^{2}}{4}s^{4},{}\xi_{3}^{''}(\omega_{0}
)=\frac{1}{bmc^{2}}[2(1+s^{2})+s^{2}],\label{SDP2}
\end{eqnarray}
\begin{eqnarray}
\xi_{2}(\omega_{0})=\frac{bmc^{2}}{5}s^{5},{}\xi_{2}^{''}(\omega_{0})=\frac{1}{bmc^{2}}[3s(1+s^{2})+s^{3}].\label{SDP3}
\end{eqnarray}
Substitution of the equations (\ref{SDP1}), (\ref{SDP2}) and
(\ref{SDP3}) in the equation (\ref{FDEN}) leads us to the required
result of the TF charge density which is compatible to
relativistic regime:
\begin{eqnarray}
n(r)&=&
\frac{m^{3}c^{3}}{\pi^{2}\hbar^{3}}\frac{1}{bmc^{2}}\{\frac{bmc^{2}}{3}s^{3}
+\frac{\pi^{2}}{6}\frac{1}{bmc^{2}}(\frac{2s^{2}+1}{s})+......\nonumber\\
&+& 4\eta
mc[\frac{bmc^{2}}{4}s^{4}+\frac{\pi^{2}}{6}\frac{1}{bmc^{2}}(2(1+s^{2})+s^{2})+.....]\nonumber\\
&+&6\eta^{2}m^{2}c^{2}[\frac{bmc^{2}}{5}s^{5}+\frac
{\pi^{2}}{6}\frac{1}{bmc^{2}}(3s(1+s^{2})+s^{3})+....]+o(\eta^{3})\}
\end{eqnarray}
Poisson's equation with the use of this TF density leads to obtain
the general form of coulomb potential in the relativistic regime
where Planck-scale correction has got incorporate with the
framework used here.
\begin{eqnarray}
\nabla^{2}\phi &=& 4\pi en(r)\simeq 4\pi
e\frac{m^{3}c^{3}}{3\pi^{2}\hbar^{3}}
[\frac{(\mu+mc^{2}+e\phi)^{2}}{m^{2}c^{4}}-1]^{\frac{3}{2}}\{1+\frac{\pi^{2}}{2(bmc^{2})^{2}}
[\frac{(\mu+mc^{2}+e\phi)^{2}}{m^{2}c^{4}}-1]^{-2}\nonumber \\
&+&4\eta
mc[\frac{3}{4}(\frac{(\mu+mc^{2}+e\phi)^{2}}{m^{2}c^{4}}-1)^{\frac{1}{2}}
+\frac{\pi^{2}}{2(bmc^{2})^{2}}[\frac{(\mu+mc^{2}+e\phi)^{2}}{m^{2}c^{4}}-1]^{-2}] \nonumber\\
&+& 6\eta^{2}m^{2}c^{2}
[\frac{3}{5}(\frac{(\mu+mc^{2}+e\phi)^{2}}{m^{2}c^{4}}-1)
+\frac{\pi^{2}}{2(bmc^{2})^{2}}[\frac{(\mu+mc^{2}+e\phi)^{2}}{m^{2}c^{4}}-1]^{-2}]\}
\end{eqnarray}
If we now adopt the new variables
\begin{equation}
\Phi=\frac{\phi +\frac{\mu}{e}}{Ze/r},~r=ax,~
a=[\frac{9\pi^{2}}{128Z}]^{\frac{1}{3}}\frac{\hbar^{2}}{me^{2}},~
\lambda \equiv
[\frac{4Z^{2}}{3\pi}]^{\frac{2}{3}}\frac{e^{4}}{\hbar^{2}c^{2}},
\end{equation}
we will reach to the final expression of relativistic TF equation
in the presence of linear-quadratic generalization:
\begin{eqnarray}
\frac{d^{2}\Phi}{dx^{2}}&=&\frac{\Phi^{\frac{3}{2}}}{\sqrt{x}}[[1+\lambda\frac{\Phi}{x}]^{\frac{3}{2}}
\{1+\rho_r\frac{x^{2}}{\Phi^{2}}[1+\lambda\frac{\Phi}{x}]^{-2}
+\frac{4}{mc}\sigma_r(\frac{3}{4}[1+\lambda\frac{\Phi}{x}]^{\frac{1}{2}}+\frac{x^{2}}{\Phi^{2}}
[1+\lambda\frac{\Phi}{x}]^{-2})\nonumber\\
&+&\frac{6}{m^{2}c^{2}}
\tilde{\sigma_r}(\frac{3}{5}[1+\lambda\frac{\Phi}{x}]+\frac{x^{2}}
{\Phi^{2}}[1+\lambda\frac{\Phi}{x}]^{-2})]\label{FE}
\end{eqnarray}
 where $\rho_r$, $\sigma_r$ and $\tilde{\sigma_r}$ have the
 following expression respectively
\begin{equation}
\rho_r= \frac{1}{8}\frac{\pi^{2}a^{2}}{s^{2}e^{2}Z^{2}},~
\sigma_r= \frac{1}{8}\frac{\pi^{2}m^{2}\eta a}{s^{2}e^{2}Z},
~\tilde{\sigma_r}=
\frac{1}{8}\frac{\pi^{2}m^{4}\eta^{2}a^{2}}{s^{2}e^{4}Z^{2}}
\end{equation}
The equation (\ref{FE}) will be equally useful in the vicinity of
Plank-scale when relativistic effect will be taken into account.
Note that in the limit $p\ll mc$, the above equation leads to TF
equation in the absence of relativistic effects:
\begin{equation}
\frac{d^{2}\Phi}{dx^{2}}=\frac{\Phi^{\frac{3}{2}}}{\sqrt{x}}\{1+\rho_r\frac{x^{2}}
{\Phi^{2}}+\frac{4}{mc}\sigma_r[\frac{3}{4}+\frac{x^{2}}{\Phi^{2}}]
+\frac{6}{m^{2}c^{2}}\tilde{\sigma_r}[\frac{3}{5}+\frac{x^{2}}{\Phi^{2}}]\}.
\end{equation}
In the limit $\eta\rightarrow 0$, or in the absence of thermal
effects, the relativistic TF equation becomes
\begin{equation}
\frac{d^{2}\Phi}{dx^{2}}=\frac{\Phi^{\frac{3}{2}}}{\sqrt{x}}[1+\lambda\frac{\Phi}{x}]^{\frac{3}{2}}
\end{equation}.
\section{Summary and Discussion}
This present paper is an extension and elaboration of the TF model
with linear-quadratic generalization to make it potent to capture
the Plank-scale effect. Initially, we consider the
non-relativistic case. It is then extended to the relativistic
regime. For the non-relativistic case screening process has been
studied with the evaluation of screening length. For the
relativistic case also the screening length is easily calculable.
The generalization done here may be useful for high-density system
as a star. How the screening process will be affected in the
quark-gluon plasma system by the quantum gravity effect can also
be studied using this modified TF equation for quark-gluon plasma
at a very high density.
\section{Acknowledgements}
AR likes to  acknowledge the facilities extended to him during his
visit to the I.U.C.A.A, Pune. He also likes to thank the Director
of Saha Institute of Nuclear Physics, Kolkata, for providing
library facilities of the Institute.

\end{document}